\documentclass[aps,%
               prl,%
               reprint,%
               showpacs,%
               superscriptaddress,%
               longbibliography,%
               nofootinbib%
               ]{revtex4-2}

\usepackage[utf8]{inputenc}
\usepackage[usenames,dvipsnames]{xcolor}
\usepackage{graphicx}
\usepackage{nicematrix}
\usepackage{amsfonts}
\usepackage{amssymb}
\usepackage{amsmath}
\usepackage{mathtools}
\usepackage{bm}
\usepackage{dsfont}
\usepackage{braket}
\usepackage{tikz}
\usepackage{txfonts}

\usepackage[pdfusetitle,%
            bookmarks=true,%
            colorlinks,%
            linkcolor=blue,%
            citecolor=blue,%
            urlcolor=blue%
            ]{hyperref}

\newcommand{\eq}[1]{Eq.\thinspace(\ref{#1})}

\newcommand{\fig}[1]{Fig.\thinspace{}\ref{#1}}

\newcommand{\subfigref}[2]{\hyperref[fig:#1]{\ref*{fig:#1}(#2)}}

\newcommand{\TUM}{\affiliation{Technical University of Munich, TUM School of Natural Sciences, Physics Department, 85748 Garching, Germany}}
\newcommand{\MCQST}{\affiliation{Munich Center for Quantum Science and Technology (MCQST), Schellingstr. 4, 80799 M{\"u}nchen, Germany}}

\begin{document}

\def\papertitle{{Entanglement Pattern Transition of Quantum States from Directed Percolation}}

\title{\papertitle}

\author{Julian Boesl} \TUM \MCQST
\author{Frank Pollmann}  \TUM \MCQST
\author{Michael Knap}  \TUM \MCQST

\date{\today}

\begin{abstract}
Changes in the entanglement structure and critical phenomena are hallmarks of quantum phase transitions. 
Here, we discuss how they appear in transitions between classes of states with distinct entanglement patterns beyond the paradigm of stable equilibrium phases of matter. Using a mapping between stochastic automata and isometric Tensor Network States (isoTNS), we construct a two-dimensional quantum state from the Domany-Kinzel automaton, which is a (1+1)D process with an absorbing phase transition in the directed percolation class. 
At the critical point of the automaton, the corresponding isoTNS hosts algebraic correlations in all spatial directions. 
The continuous parent Hamiltonian of this state has a degenerate ground state manifold.
It consists of a product state (the absorbing state) and a second state that undergoes a transition from pairwise entanglement between distant regions, similarly to the $W$ state, to a state with trivial entanglement. 
Our results demonstrate how the correspondence between isoTNS and classical stochastic evolution can be used to probe the Hilbert space structure beyond stable ground state manifolds.
\end{abstract}

\maketitle

\textbf{\emph{Introduction.---}} Quantum phases of matter can be characterized by the entanglement pattern of their ground states~\cite{vidal_entanglement2003, zeng2019quantum, jiang2012identifying}. Topologically ordered states feature topological entanglement entropy, which contains information about their anyonic excitations~\cite{kitaev_topological2006, levin_detecting2006, chen_tensor2010, papanikoaou_topological2007, castelnovo_quantum2008}; many of them are captured by string-net models, which provide simple fixed point wavefunctions~\cite{levin_string2005, gu_tensor2009, buerschaper_explicit2009}. Similarly, symmetry-breaking and symmetry-protected topological (SPT) states have multipartite entanglement~\cite{chen2015discontinuity, zeng_gapped2015, zeng2016topological} and are represented by states such as the GHZ state~\cite{greenberger1990bell} or the cluster state~\cite{briegel_persistent2001} for the transverse field Ising model and the cluster model at zero field, respectively. These examples are gapped phases of matter, as their ground state degeneracy is stable to local (symmetry-preserving) perturbations~\cite{bravyi2011short, hastings_quasiadiabatic2005, chen_classification2011}. However, the space of non-trivially entangled states includes more fragile states~\cite{dur_three2000, verstraete_four2002, horodecki_entanglement2009}, which do not lead to many-body phases in the thermodynamic limit. Still, one can investigate their parent Hamiltonians: Recent work has explored $W$ and Dicke states as ground states and many-body scars~\cite{gioia2024wstate, gioia2025distinct, odea2026locality}.

Tensor Network States (TNS) are a powerful tool to study quantum phases~\cite{cirac_matrix2021, verstraete_criticality2006, perezgarcia_peps2007, schuch_peps2010}; for instance, they provide continuous wavefunction paths between the aforementioned 1D fixed points~\cite{wolf_quantum2006, jones_skeleton2021, camp_matrix2025, hallam2026spectral}. A subclass of higher-dimensional TNS are isometric TNS (isoTNS), which have favorable contraction properties and guarantee efficient preparation protocols~\cite{zaletel_isometric2020, soejima_isometric2020, slattery_quantum_2021, wei_sequential2022}. Certain isoTNS can be constructed from stochastic classical automata; update rules conserving a parity correspond to abelian string-nets, providing exact phase transitions between different topological orders~\cite{liu_simulating2024, boesl_quantum2025, boesl2025skeleton}. However, stochastic automata also feature genuine non-equilibrium transitions such as absorbing phase transitions, where the system cannot escape certain states~\cite{hinrichsen_nonequilibrium2000}. The most common universality class is directed percolation (DP), describing the proliferation of clusters which may die out, but never be spontaneously created~\cite{odor_universality2004}. At low survival probabilities all states eventually reach the empty state, while at high survival probabilities a fluctuating steady state with a finite density of active sites appears. The automaton-to-isoTNS mapping raises the question how these transitions manifest in the related quantum states.
\begin{figure}
\centering
    \includegraphics[width=.9\columnwidth]{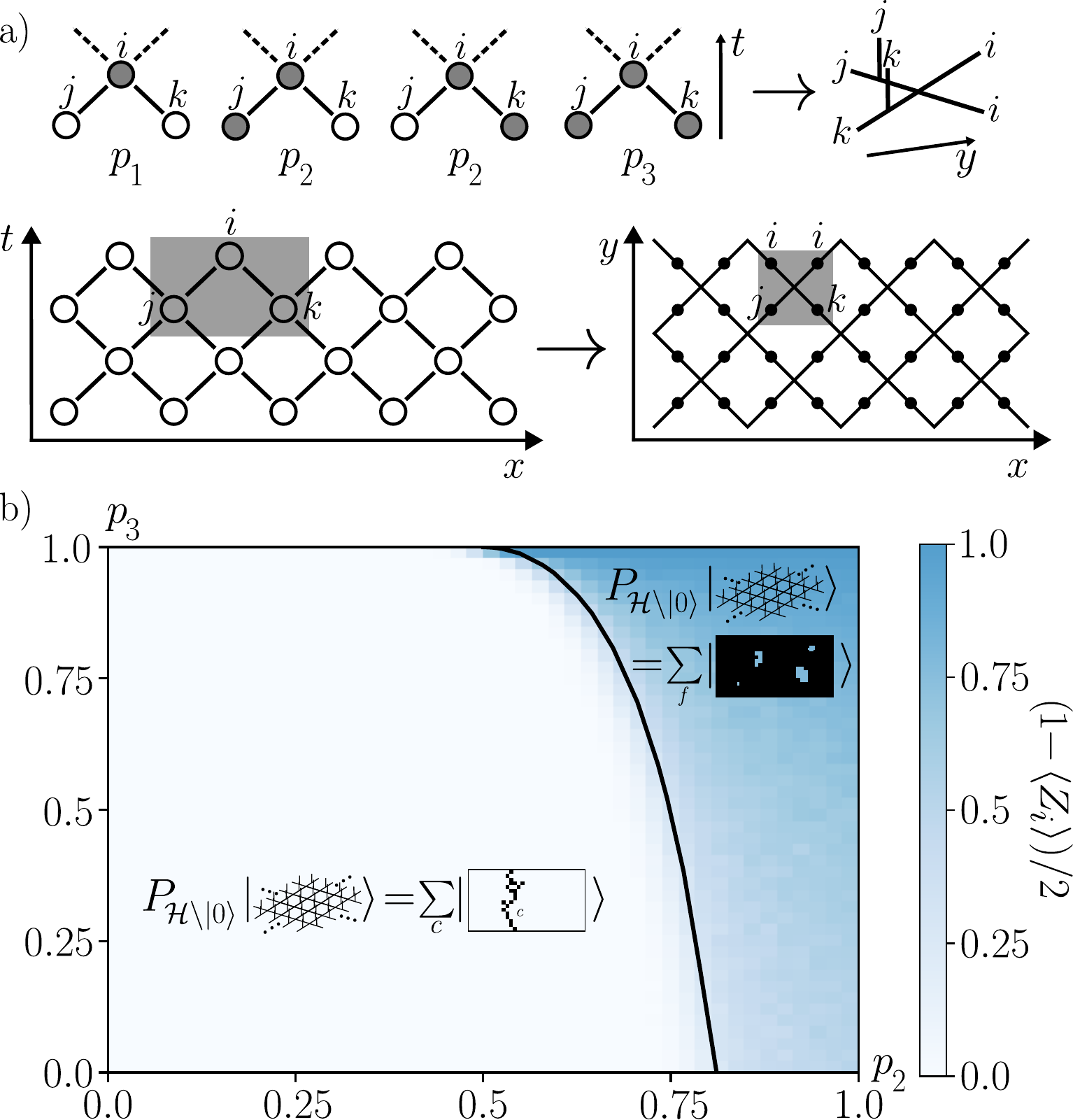}
    \caption{\textbf{The Domany-Kinzel (DK) automaton.} a) In the DK automaton, even ($i$) and odd ($j$ and $k$) sites are updated alternatingly. White/gray circles represent empty/active sites ($0/1$); the probabilities of an empty site at $i$ are $P(0\vert jk) = 1-P(1\vert jk)$. By adding physical legs, the automaton is mapped to a quantum state; time $t$ becomes a spatial direction $y$. b) The phase diagram of the DK automaton for $p_1 = 0$. At low $p_{2/3}$, the automaton is in the absorbing phase; the order parameter $\overline{n}_i = (1-\langle Z_i \rangle)/2$ is zero. In the active phase at high $p_{2/3}$, $\overline{n}_i$ is finite. The transition line (black) is taken from~\cite{henkel2008nonequilibrium}. For periodic boundaries, the ground state manifold of the parent Hamiltonian is degenerate and hosts a state with different entanglement depending on the phase (insets): A delocalized system-spanning cluster of active sites with pairwise entanglement between distant regions (absorbing phase) and a trivially entangled state (active phase).
    }
    \label{fig:MapPD}
\end{figure}

In this work, we discuss how the two perspectives combine. Starting from a (1+1)D model in the DP class, the Domany-Kinzel automaton~\cite{DomanyKinzel, harada_entropy2019, chen2026localreversibility}, we construct the corresponding 2D isoTNS wavefunction and its local parent Hamiltonian, that is continuous across the transition. At the transition, the state supports algebraic correlations in all directions, a feature hitherto unseen in isoTNS. The quantum perspective on the transition becomes more apparent on periodic lattices: The ground state manifold of the parent Hamiltonian is degenerate, featuring the empty state and a second ground state. In the active phase of the automaton, it represents the fluctuating steady state and is trivially entangled, while in the absorbing phase a delocalized cluster spans the entire time-like direction, whose width in the space-like direction diverges at the critical point. Arbitrarily distant regions are pairwise entangled, reminiscent of the $W$ state. The model hosts a transition between entanglement patterns which do not define stable phases of matter, and demonstrates how ideas from non-equilibrium physics may aid in exploring the Hilbert space structure of non-trivial states beyond many-body ground states.

\textbf{\emph{Domany-Kinzel automaton.---}} Classical non-equilibrium physics allows for phase transitions without static counterparts, which are realized in models violating detailed balance~\cite{odor_universality2004}. One example are absorbing phase transitions, which feature at least one state which the system cannot escape once it has entered. It has been conjectured that under some mild additional conditions, a generic absorbing phase transition with a single absorbing state is in the universality class of directed percolation (DP)~\cite{hinrichsen_nonequilibrium2000}. In DP, a site is either ``empty" or ``active"; active sites may spread further with some probability $p$, or otherwise die out, whereas isolated empty sites remain empty. The fully empty system thus is an absorbing state. The system may undergo a phase transition at some critical $p_c$, where a fluctuating steady state is stabilized. This state corresponds to an active cluster spanning the entire system; its density $n$ of active sites is an order parameter above $p_c$, scaling as $n \sim \vert p - p_c\vert^{\beta}$ close to the critical point. At $p = p_c$, this cluster has zero density and a fractal structure, corresponding to the divergence of all length scales. There are two correlation lengths $\xi_{\parallel/\perp}$ in time and space, which diverge with different exponents $\nu_{\parallel/\perp}$, $\xi_{\parallel/\perp} \sim \vert p - p_c\vert^{\nu_{\parallel/\perp}}$.

In one dimension, a simple example of DP is the Domany-Kinzel (DK) automaton~\cite{DomanyKinzel}. Even and odd sites on a chain are updated at alternating time steps, a site $i$ being updated depending on its nearest neighbors $j$ and $k$; we can thus visualize the process on a diagonal square lattice. Assuming inversion symmetry and labeling empty/active sites as 0/1, this leaves three parameters defining the conditional probabilities $P(i\vert jk)$,
\begin{align}
    P(1\vert00) = p_1, \;P(1\vert01) = P(1\vert10) = p_2, \; P(1\vert11) = p_3,
    \label{eq:DKDefinition}
\end{align}
and $P(0\vert jk) = 1- P(1\vert jk)$ (see \fig{fig:MapPD}a) for a graphical representation). For $p_1= 0$, the empty state is absorbing, allowing DP physics to govern the system. Numerical studies confirm the phase diagram shown in \fig{fig:MapPD}b), with an absorbing phase at low $p_{2/3}$ and an active phase at high values~\cite{hinrichsen_nonequilibrium2000, henkel2008nonequilibrium}. The critical line in the $p_2-p_3$ plane lies in the DP universality class except for the point $p_2 = 0.5, p_3 = 1$, where an additional symmetry modifies the transition~\cite{DomanyKinzel, odor_universality2004}.

\textbf{\emph{Automaton-to-isoTNS mapping.---}} A classical stochastic process in $d$ spatial dimensions can be mapped to a $d +1$ dimensional quantum state, where time becomes an additional spatial direction. The quantum state $\ket \Psi$ is a superposition of all possible trajectories $\alpha$ as $\ket \Psi = \sum_\alpha \sqrt {p_\alpha} \ket \alpha$, $p_\alpha$ being the probability of the trajectory $\alpha$~\cite{boesl2025skeleton,gopalakrishnan2025push, zhang_sequential2026}.

For stochastic cellular automata with a brickwork structure such as the DK automaton, this mapping is local, yielding a tensor network state (TNS)~\cite{liu_simulating2024, yu_dual2024, boesl_quantum2025, boesl2025skeleton}: The local update rule $P(i\vert jk)$ can be associated with a tensor $T$, whose virtual legs are split in two sets in opposite directions, one carrying the information about the states $j$ and $k$ in the preceding layer, while the other legs feed forward information about the current local state $i$. For each incoming virtual degree of freedom, an additional physical leg is added which is locked with its virtual counterpart. The DK automaton thus determines a 2D tensor $T_{\text{DK}}$ with bond dimension $\chi = 2$ as
\begin{equation}
    \left(T_{\text{DK}}\right)^{jk}_{abcd} = \sqrt {P(i\vert jk)} \delta_{a,j}\delta_{b,k}\delta_{c,i}\delta_{d,i},
    \label{eq:DKTensor}
\end{equation}
which is visualized in \fig{fig:MapPD}a). The full state $\ket{\text{DK}}$ is a tensor contraction over all virtual legs,

\begin{equation}
    \ket{\text{DK}} = \sum_{j_1, \cdots, j_k} \text{tTr}\left( \left\{ T^{j_1j_2}, \cdots, T^{j_{k-1}j_k} \right\} \right) \ket{j_1\cdots j_k},
    \label{eq:TensorWF}
\end{equation}
leading to a state on a square lattice with one qubit per edge. The conservation of probability, $\sum_i P(i \vert jk) = 1 \;\forall j,k $, in the update rule corresponds to an isometry condition on the local tensor, $\sum_{j,k,c,d} \left(T_{\text{DK}}\right)^{jk}_{abcd} \left(\left(T_{\text{DK}}\right)^{jk}_{a^\prime b^\prime cd}\right)^* = \delta_{a, a^\prime} \delta_{b, b^\prime}$. Such isometric TNS (isoTNS) form an expressive subclass of TNS which includes string-net fixed points and a set of finite correlation length deformations~\cite{soejima_isometric2020, boesl2025skeleton}. They are of particular interest as they allow for sequential preparation circuits which scale in linear system size. This becomes apparent for stochastic automata states, as the update rule is directly implemented by a unitary gate~\cite{slattery_quantum_2021, wei_sequential2022}. The descriptive range of isoTNS is still open; a salient question concerns the types of criticality they can support~\cite{haag_typical_2023, malz_computational_2024}.

The frustration-free parent Hamiltonian $H_{\text{DK}}$ for the state $\ket{\text{DK}}$ in \eq{eq:TensorWF} is a sum of 8-qubit projectors. A detailed definition of this Hamiltonian $H_{\text{DK}}$ is presented in Appendix A. It is continuous for all values $(p_1, p_2, p_3)$ with $p_{2/3} \neq  0$, in particular, including the interior of the $p_1 = 0$ plane where the absorbing phase transition arises; Fig.~\ref{fig:MapPD}b).

\textbf{\emph{Correlation functions and critical isoTNS.---}} For open boundaries, we evaluate the expectation value of any diagonal operator by sampling the associated observable in the corresponding stochastic automaton with an initial distribution given by the boundary state (more generally, a general operator can be evaluated by explicit construction of a diagonal operator with the same expectation value on the TNS~\cite{boesl2025skeleton}). We define the two directions $x$ (space) and $y$ (time); i.e., they are rotated by 45 degrees compared to the standard lattice vectors. The expectation value of the $Z$ operator (with $Z\ket{0/1} = \pm\ket{0/1}$) on a site $i$ with coordinates $x$ and $y$ is $\langle Z_{x,y} \rangle = 1-2\overline{n}_x(t = y)$. We define a correlation function
\begin{equation}
    C^i(j) = \frac{1}{4}  \langle (1-Z_i)(1-Z_j)\rangle = \overline{n_i n_j}, 
    \label{eq:Corr}
\end{equation}
which corresponds to a correlation of densities in the stochastic automaton. In the bulk of a thermodynamically large system, the correlations are bounded by the density of active sites $C^i(j) \leq n$, i.e., they vanish outside of the active phase~\cite{dickman2005quasi}. This implies that they are zero at the phase transition as there the critical cluster has zero density. It is pertinent to consider a normalized correlation function $C_{\text{norm}}^i(j)$ in the bulk of a finite-size system~\cite{masaoka_rigorous2025} defined as
\begin{equation}
    C^i_{\text{norm}}(j) = \frac{\langle (1-Z_i)(1-Z_j)\rangle}{2(1- \langle Z_i\rangle)} = \frac{\overline{n_i n_j}}{\overline{n}_i}.
    \label{eq:CorrNorm}
\end{equation}
This function is finite for any system size, even in the absorbing phase, as $C^i_{\text{norm}}(i) = 1$; importantly, this expression remains finite when approaching the thermodynamic limit.

\begin{figure}
\centering
    \includegraphics[width=.95\columnwidth]{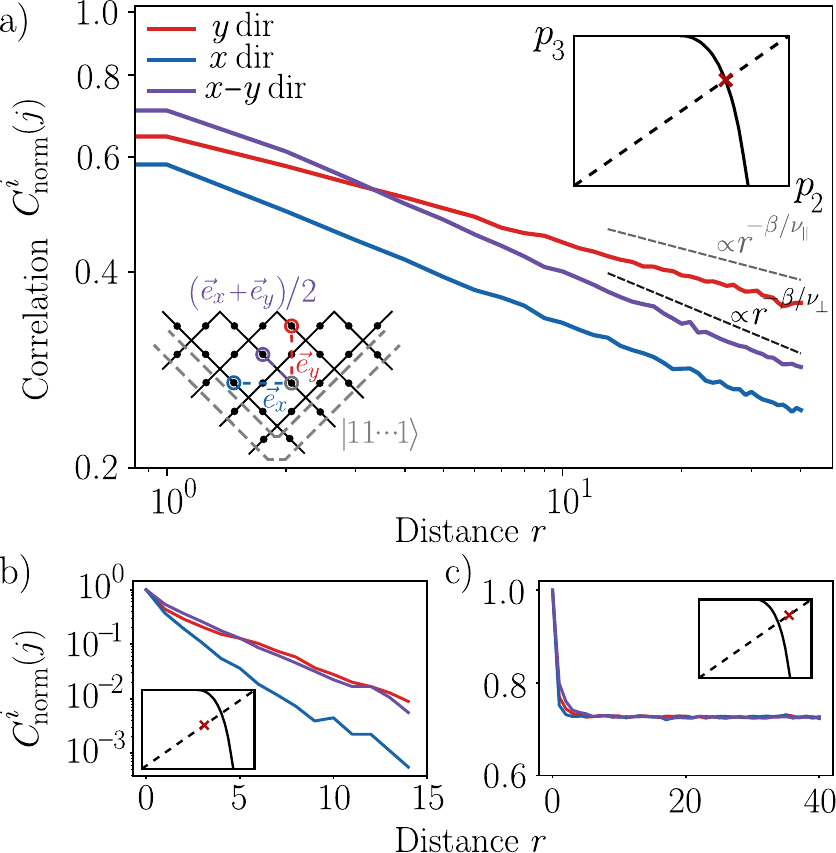}
    \caption{\textbf{Normalized correlations.} The normalized correlations $C^i_\text{norm}(j)$ in the directions $y$ (red), $x$ (blue) and $x-y$ (purple), evaluated in an open boundary system with fully active boundaries for $p_2 = p_3 = p$. a) Correlations on the critical line $p = 0.7055$. All directions feature algebraic scaling with exponent $\beta/\nu_\parallel$ in $y$ direction and $\beta/\nu_\perp$ in all other directions. b) Correlations in the absorbing phase, $p = 0.55$ decay exponentially. c) Correlations in the active phase, $p = 0.8$ saturate to a finite value, corresponding to the density $n$ of active sites in the steady state. The linear system size is $L = 3000$ in a) and c), and $L= 100$ in b).
    }
    \label{fig:DKCorr}
\end{figure}

In \fig{fig:DKCorr}, we show numerical results for $C^i_{\text{norm}}(j)$. Two edges of a square system meeting at a corner are initialized in the state $\ket {11\cdots 1}$, corresponding to a boundary of active states (dashed box). The site $i$ is chosen to be in the middle of the lattice to reduce boundary effects. We show correlations in the $y$ (time-like), $x$ (space-like) and $x-y$ (diagonal) directions. We focus on the line $p_2 = p_3 = p$, representing site directed percolation, where accurate predictions for the critical point $p_c \approx 0.7055$ exist~\cite{hinrichsen_nonequilibrium2000}. In the top panel, we show the results for the three directions at $p_c$. All directions are governed by the critical exponent $\nu_\perp$ of the spatial direction $x$, expect for the purely time-like direction $y$: $C^i_{\text{norm}}(j) \sim \vert i- j\vert^{-\beta/\nu_\parallel}$ in $y$ direction and $C^i_{\text{norm}}(j) \sim \vert i- j\vert^{-\beta/\nu_\perp}$ in all other directions, with  $\beta/\nu_\parallel \approx 0.1595$ and $\beta/\nu_\perp \approx 0.2521$, respectively.

In the absorbing phase $p < p_c$, the correlations decay exponentially; they probe the extent of a cluster of size $\xi_{\parallel}$ in $y$ direction and $\xi_{\perp}$ in $x$ direction provided site $i$ is active, which becomes unlikely away from the boundary as $\overline{n}_i$ decays exponentially with this distance. By contrast, in the active phase $p > p_c$ the correlations saturate to the density $n$ of active sites.

Note that the state at $p_2 = p_3=p_c$ (and in general on the critical line) is a first example of an isoTNS which hosts algebraic correlations in all spatial directions; earlier examples exhibit only one critical direction~\cite{liu_simulating2024, boesl_quantum2025, boesl2025skeleton}. The definition of the correlation function arises from the associated transition, whose quantum nature we discuss in the following section.

\textbf{\emph{Directed percolation as transition between entanglement patterns.---}} For open boundaries, the physics of the quantum state can be interpreted directly in the framework of the classical stochastic automaton. However, the tensor network state can also be put on a lattice which is periodic not only in the $x$ direction (corresponding to spatial dimension of the automaton), but also in the $y$ direction (corresponding to time in the automaton). This change of topology has important ramifications for the quantum system: In the stochastic automaton at $p_1 = 0$, active sites which are spontaneously created from two empty sites are disallowed. In the quantum system, we call such a local configuration a ``defect"; the parent Hamiltonian $H_\text{DK}$ exhibits a $U(1)$ conservation law of the number $N_D$ of defects, as it cannot create or annihilate them, splitting the Hilbert space into sectors labeled by $N_D$; see Appendix A.

By construction, the ground state manifold of $H_\text{DK}$ lies in the $N_D = 0$ sector, and $\ket{\text{DK}}$ belongs to it. The completely empty state $\ket{\text{vac}} = \ket{00\cdots 0}$ is also a ground state, as it is annihilated by all projectors. This follows as it cannot be connected to any other zero-defect state by a local operation; it can only be connected to other configurational basis states without defects via the creation of a system-spanning cluster wrapping around the periodic $y$ direction.

As both $\ket{\text{DK}}$ and $\ket{\text{vac}}$ are ground states and $\langle\text{DK}\ket{\text{vac}} \neq 1$ for all system sizes, the ground state manifold is degenerate, including a state $\ket{\text{GS}}$ orthogonal to $\ket{\text{vac}}$ defined as

\begin{equation}
    \ket{\text{GS}} = \frac{P_{\mathcal{H}/\ket{\text{vac}}} \ket{\text{DK}}}{\vert\vert P_{\mathcal{H}/\ket{\text{vac}}} \ket{\text{DK}} \vert\vert},
    \label{eq:GroundStateNonTriv}
\end{equation}
where $P_{\mathcal{H}/\ket{\text{vac}}} = 1-\ket{\text{vac}}\bra{\text{vac}}$ projects out the trivial state. Although $\lim_{L_x,L_y \rightarrow \infty}\ket{\text{DK}} = \ket{\text{vac}}$ in the absorbing phase, the state $\ket{\text{GS}}$ also remains a ground state of $H_{\text{DK}}$.

To understand the nature of this second ground state, we consider two contracted rows in $x$ direction of the tensor $T_{\text{DK}}$, assuming $L_x \rightarrow \infty$. This object $\mathbb{T}$ corresponds to the transfer matrix of the DK automaton, with the physical legs storing the information of the even sites $i$ at time $t$ and the odd sites $i + \frac{1}{2}$ at time $t +1$. We split this object as $\mathbb{T} = P_0\mathbb{T} + (1-P_0) \mathbb{T}$, where the operator $P_0$ projects the outgoing virtual legs to the all-zero configuration. $(1-P_0) \mathbb{T}$ represents the DK automaton conditioned on the outcome that the absorbing state is not entered; in particular, as $(1-P_0) \mathbb{T}P_0 = 0$, for periodic $y$ direction $\ket{\text{DK}} = (P_0\mathbb{T})^{Ly} + \bigl((1-P_0) \mathbb{T}\bigr)^{L_y}$, where the last and the first instance of the transfer matrix are contracted so that no virtual legs remain. $(P_0\mathbb{T})^{Ly} \propto \ket{\text{vac}}$ is the empty state, while $\bigl((1-P_0) \mathbb{T}\bigr)^{L_y} \propto \ket{\text{GS}}$ contains all contributions with at least one cluster spanning the $y$ direction.

\begin{figure}
\centering
    \includegraphics[width=.75\columnwidth]{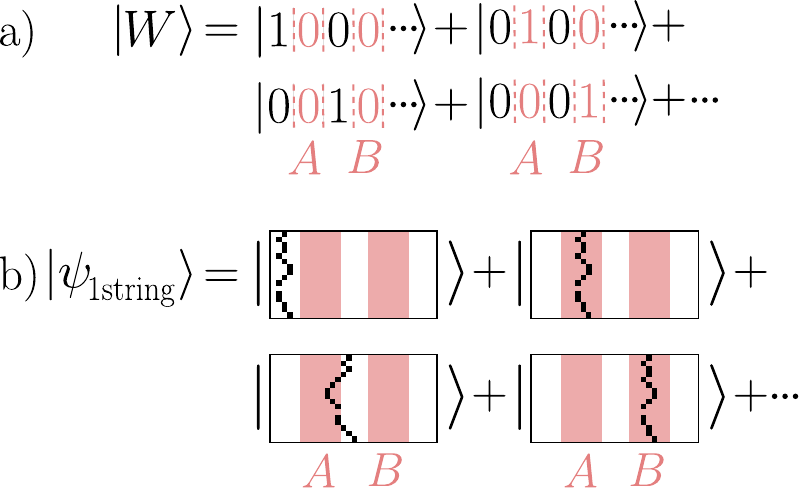}
    \caption{ \textbf{Pairwise entanglement throughout the absorbing phase.} a) The $L$-qubit $W$ state $\ket W = \frac{1}{\sqrt{L}} \left( \sum_i^L X_i \right) \ket{00 \cdots 0}$ features pairwise entanglement between any two qubits $A$ and $B$; their reduced density matrix $\rho_{AB}$ has a negativity $\mathcal{N}(\rho_{AB}) = \frac{1}{L(L-2)}$. b) The ground state $\ket{\text{GS}}$ Eq.~\eqref{eq:GroundStateNonTriv} in an $L_x \times L_y$ system deep in the absorbing phase $p_1 = 0, p_2 = p_3 = \varepsilon \ll 1 $ approaches the state $\ket{\psi_{\text{1string}}}$, a superposition of all ways a single string of 1's can wrap around the periodic $y$ direction on top of the empty state. Defining two separated regions $A$ and $B$ with finite extent $\ell$ in $x$ direction, the string can either lie completely outside, completely inside or partially in and out of $A$ and $B$; for $\ell > L_y$, the reduced density matrix $\rho_{AB}$ has a negativity $\mathcal{N}(\rho_{AB}) \geq \frac{1}{L_x(L_x-2)}$, representing a similar entanglement pattern as the $W$ state.
    }
    \label{fig:WComparison}
\end{figure}

For values of $p_{2/3}$ in the absorbing phase, the empty state is the only eigenstate of the DK automaton with eigenvalue 1, implying the contribution of $\bigl((1-P_0) \mathbb{T}\bigr)^{L_y}$ to $\ket{\text{DK}}$ vanishes in the thermodynamic limit as expected. As $L_y \rightarrow \infty$, $\ket{\text{GS}}$ is dominated by the eigenstate of the conditioned DK automaton with the highest eigenvalue. Deep in the absorbing phase $p_2 = p_3 = \varepsilon \ll 1$, this eigenstate approaches $\ket{\psi} = \frac{1}{\sqrt{2L_x}} \left( \sum_i X_{i,t}(X_{i-\frac{1}{2}, t+1} + X_{i+\frac{1}{2}, t+1}) \right)\ket{00\cdots 0}$, i.e. a superposition of a single minimal-size cluster. Therefore, the state $\ket{\text{GS}}$ is the superposition $\ket{\psi_{\text{1string}}}$ of all ways a single string of active sites can wrap around the $y$ direction. This ground state has non-trivial entanglement: If we take two regions $A$ and $B$ spanning the entire $y$ direction with a finite extent $\ell > L_y$, these regions will be pairwise entangled even when arbitrarily far apart. Indeed, their reduced density matrix $\rho_{AB}$ features a non-vanishing negativity
\begin{equation}
    \mathcal{N}(\rho_{AB}) \geq \frac{1}{L_x(L_x -2)},
\end{equation}
which does not depend on the distance $d$ between $A$ and $B$ (see Appendix B). This pairwise entanglement resembles the $W$ state $\ket W = \frac{1}{\sqrt{L}} \left(\sum^L_i X_i \right)\ket{00 \cdots 0}$, which features entanglement between every pair of qubits, making it more stable to local loss compared to the multipartite entanglement of the GHZ state~\cite{dur_three2000}; see \fig{fig:WComparison} for a graphical comparison. As $p_{2/3}$ are increased, the delocalized cluster in $\ket{\text{GS}}$ broadens in $x$ direction on a length scale $\xi_\perp$. Still, the global entanglement structure remains unchanged, as there is still pairwise entanglement between $A$ and $B$ provided $\ell, d \gg L_y, \xi_\perp$ and $L_y \gg \xi_\parallel$, as higher-order contributions are suppressed. In particular, if the system apart from $A$ and $B$ is measured in the all-zero state, the cluster will be in a Bell state between $A$ and $B$.

This entanglement pattern changes at the critical line of the DK automaton: The width of the delocalized string diverges and gives rise to the non-trivial steady state, corresponding to an eigenstate of $(1-P_0)\mathbb{T}$ with eigenvalue 1. In the active phase, for $L_y \gg \xi_\parallel$, the second ground state $\ket {\text{GS}}$ is given by this steady state. As it is a global cluster of active sites of density $n$, with droplets of empty sites of size $\xi_{\parallel/\perp}$ in $y/x$ direction, the two regions $A$ and $B$ are not entangled if their distance $d > \xi_\perp$ exceeds the correlation length. In this regime, $\ket{\text{GS}}$ is thus a paramagnetic state with trivial short-range entanglement on length scales $\xi_{\perp/\parallel}$.

The presence of the transition between the two classes of states ($W$-like to trivial) intimately depends on the presence of the $U(1)$ conservation law of defects: If we allow for $p_1 \neq 0$, $\ket{\text{vac}}$ is not fully disconnected anymore and ceases to be an exact ground state. Accordingly, $(1-P_0)\mathbb{T}P_0 \neq 0$ and $\ket{\text{DK}}$ becomes the unique, trivially entangled ground state of $H_{\text{DK}}$, given by the steady state of the DK automaton. Other classical models with similar conservation laws do not necessarily lead to an equivalent transition: In the supplement~\cite{supp}, we show how for instance a quantum state constructed from the Kasteleyn model~\cite{kasteleyn1963dimer} leads to a parent Hamiltonian without this transition, due to additional symmetry constraints.

This fragility of the ground state manifold demonstrates that the two classes of states do not represent stable gapped quantum phases of matter; rather, a generic perturbation will immediately lift the degeneracy, even if respects the $U(1)$ defect number symmetry; an example would be an on-site $Z$ field. Recent work has shown that in the thermodynamic limit, the $W$ state can only appear as the ground state of a Hamiltonian in the presence of an additional empty state $\ket{\text{vac}}$~\cite{gioia2024wstate, gioia2025distinct}. This condition applies to our model as well, 
in which the size of the single cluster diverges as the phase transition is approached and the state becomes locally distinguishable from $\ket{\text{vac}}$, by $Z$ measurements for example.

\textbf{\emph{Outlook.---}} In this work, we have introduced a quantum state obtained from the Domany-Kinzel automaton through the automaton-to-isoTNS mapping. In the quantum system, the non-equilibrium transition reappears as a transition between different entanglement patterns, with the absorbing phase exhibiting pairwise entanglement between regions of arbitrary distance, similar to the $W$ state, and the active phase being trivially entangled. The restriction to isoTNS is in fact not necessary for the physics at play; non-isometric deformations should exhibit similar physics as long as the degeneracy of the ground state manifold is left unchanged. In particular, the tensor derived here is not injective, which suggests that it cannot be the unique ground state of any parent Hamiltonian~\cite{perezgarcia_peps2007}. An interesting question is whether it is possible to derive general conditions under which this transition has to persist; this would allow for the definition of a connected class of ``$W$-like" entangled states, in the same vein as symmetries allow the distinction of different symmetry-broken phases. The vacuum state being an additional ground state seems to be of importance for the parent Hamiltonians of such states, which may be related to a virtual symmetry of the transition matrix $\mathbb{T}$. Importantly, while the $U(1)$ conservation law of defect sites is necessary to our construction, it is not sufficient to ensure this ground state degeneracy.

Furthermore, it seems promising to consider a similar approach to other states with interesting entanglement structures which are encountered in quantum information theory. The $W$ state is one example in the more general class of Dicke states~\cite{dicke_coherence1954}, and more generally there are many examples of states with non-equivalent types of entanglement~\cite{dur_three2000, verstraete_four2002, horodecki_entanglement2009}. Normally, these states are considered as isolated points in the Hilbert space; finding ways to tune them in their respective class of entanglement and across transitions enables exploring the structure of the Hilbert space beyond conventional ground states of quantum many-body systems.

Coming from the other side of the mapping, there is an entire zoo of classical stochastic dynamics that can be mapped to quantum states by our strategy. Already for the case of absorbing phase transitions, different universality classes exist in the presence of symmetries or multiple absorbing states~\cite{hinrichsen_nonequilibrium2000, odor_universality2004}, which can be translated to the quantum language following the provided example; in the supplement~\cite{supp}, we show how 2D bond directed percolation maps to a 3D quantum state. Beyond absorbing phase transitions, there are automata serving as stable memory such as Toom's rule~\cite{toomsrule, lake2025squeezingcodes} or Gacs' automaton~\cite{gacs2001reliable}, or exhibiting self-organized criticality~\cite{bak_selforganized1987}. From these examples, it is apparent that non-equilibrium physics can serve as a guiding principle to discover non-trivial quantum states beyond our current understanding.

\textit{\textbf{Acknowledgments.---}} We thank Sebastian Diehl, Sarang Gopalakrishnan, Haye Hinrichsen, Yu-Jie Liu, Tibor Rakovszky, and Yizhi You  for insightful discussions. We acknowledge support from the Deutsche Forschungsgemeinschaft (DFG, German Research Foundation) under Germany’s Excellence Strategy–EXC–2111–390814868, TRR 360 – 492547816 and DFG grants No. KN1254/1-2, KN1254/2-1, the European Union (grant agreement No 101169765), as well as the Munich Quantum Valley, which is supported by the Bavarian state government with funds from the Hightech Agenda Bayern Plus.

\textit{\textbf{Data availability.---}}Numerical codes are available upon reasonable request on Zenodo~\cite{zenodo}.

\section*{End Matter}

\section{Appendix A: Definition of parent Hamiltonian}
\begin{figure}
\centering
    \includegraphics[width=.65\columnwidth]{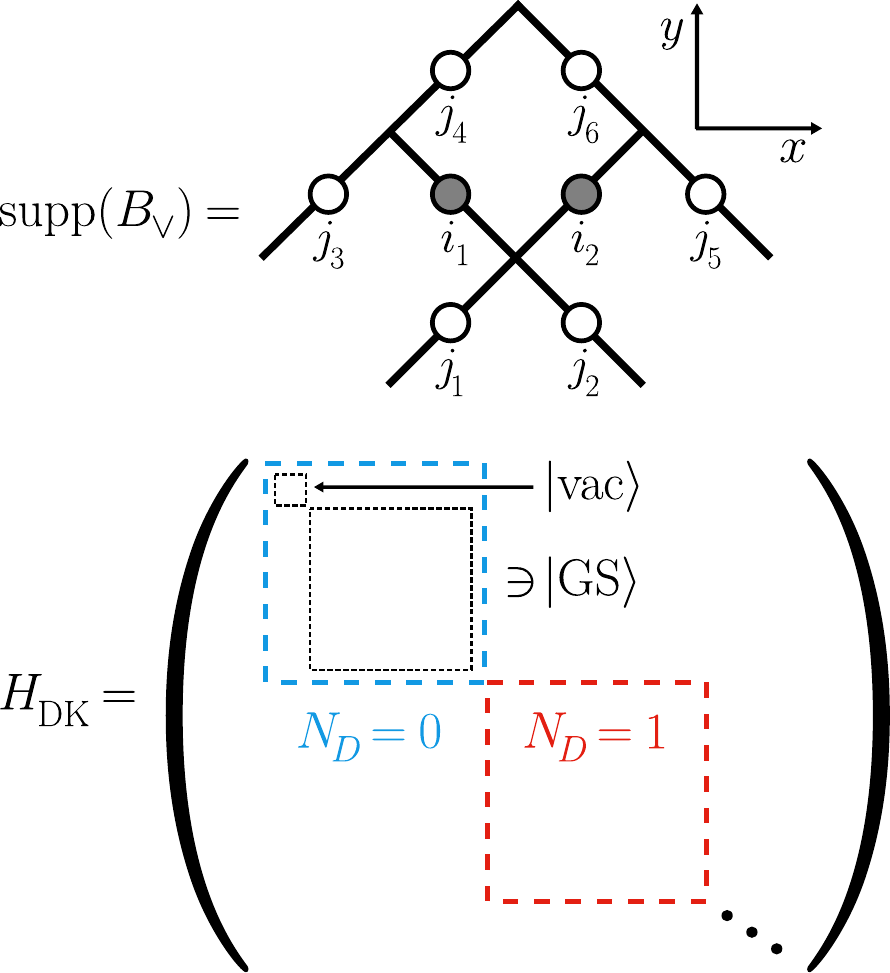}
    \caption{\textbf{Parent Hamiltonian $H_{\text{DK}}$.} Top: The support of the projector $B_\vee$. The outer qubits $j_1, \dots j_6$ are acted on only diagonally in the $Z$ basis, while the inner qubits $i_1, i_2$ are projected onto a superposition $\alpha \ket{00} + \beta \ket{11}$; the coefficients depend on the environment. Bottom: The structure of the Hamiltonian $H_{\text{DK}}$ in the DP plane $p_1 = 0$. The Hilbert space splits up into sectors labeled by the defect number $N_D$. The ground state manifold lies in the sector $N_D = 0$. As the state $\ket{\text{vac}} = \ket{00\cdots 0}$ cannot be connected to any other zero-defect state, it is as an additional ground state along the non-trivial state $\ket{\text{GS}}$.
    }
    \label{fig:ParentH}
\end{figure}

In this section, we define the parent Hamiltonian $H_\text{DK}$ of the isoTNS state $\ket{\text{DK}}$ Eq.~(\ref{eq:TensorWF}) where the local tensors Eq.~(\ref{eq:DKTensor}) are parametrized by $p_1, p_2, p_3$. It consists of local projectors and is frustration-free, i.e., the ground state $\ket{\text{DK}}$ minimizes all projectors at the same time. The model lives on a square lattice with qubits on edges. Considering periodic boundary conditions, the Hamiltonian is defined as
\begin{equation}
    H_{\text{DK}} =  \sum_\vee B_\vee.
    \label{eq:ParentHApp}
\end{equation}
For every vertex of the lattice, the symbol $\vee$ represents the pair of qubits on the adjacent edges above the vertex; see qubits $i_1$ and $i_2$ in \fig{fig:ParentH}, top. In the state $\ket{\text{DK}}$, the two qubits are always in the same $Z$ basis state. The projector $B_\vee$ has a support of 8 qubits around the vertex including the pair of qubits in $\vee$ (see \fig{fig:ParentH}, top). In the $Z$ basis, it acts off-diagonally only on the two locked qubits $i_{1/2}$, which it projects onto a superposition $\alpha \ket{00} + \beta \ket{11}$, where $\alpha, \beta$ are determined by the environment $j_1,\dots j_6$; accordingly, these six qubits are acted on diagonally. Explicitly, we can write this as
\begin{equation}
    B_\vee = \sum_{j_1,\dots,j_6} \Bigl(1 - \ket{i_{j_1\cdots j_6}}\bra{i_{j_1\cdots j_6}}\Bigr) \Bigl(1 - \ket{{j_1\cdots j_6}}\bra{{j_1\cdots j_6}}\Bigr),
    \label{eq:ProjectorDef}
\end{equation}
where $\ket{{j_1\cdots j_6}}$ is a $Z$ basis state of the six surrounding qubits and $\ket{i_{j_1\cdots j_6}}$ is the chosen state on the central qubits $i_{1/2}$. If all probabilities $p_1, p_2, p_3 \neq 0,1$, the number of non-zero entries in $T_\text{DK}$ is maximal as no transition probability in the stochastic automaton vanishes; in this case, the state $\ket{i_{j_1\cdots j_6}}$ is defined as
\begin{align}
    \ket{i_{j_1\cdots j_6}} \propto &  \sqrt{P(0\vert j_1 j_2) P(j_4\vert 0 j_3) P(j_6 \vert j_5 0) }\ket {00} \nonumber \\ & +\sqrt{P(1\vert j_1 j_2) P(j_4\vert 1 j_3) P(j_6 \vert j_5 1) }\ket {11}
    \label{eq:BStateDef}
\end{align}
and the unique ground state is given by $\ket{\text{DK}}$.

For $p_1 = 0$, there are some environments for which the state $\ket{i_{j_1\cdots j_6}}$ is not well-defined as both coefficients in \eq{eq:BStateDef} vanish. Nevertheless, we can still define the projector uniquely in the case of $p_{2/3} \neq 0,1$ by taking the limit $p_1 \rightarrow 0$ of the state $\ket{i_{j_1\cdots j_6}}$. In the stochastic automaton, $p_1 = 0$ implies there are no spontaneously created active sites in the time evolution as $P(1\vert00) = 0$. In the TNS, this prohibits the corresponding local configuration around one vertex, i.e., the configuration $i_1 = i_2 = 1$ if $j_1 = j_2 = 0$ in \fig{fig:ParentH}, top, which we therefore call ``defect." Accordingly, if the two states $\ket{j_1 \cdots j_6}\ket{00}$ and $\ket{j_1 \cdots j_6}\ket{11}$ have a different number of defects, the operator $B_\vee$ projects to the state with fewer defects. This Hamiltonian remains continuous if we change the parameters $p_1, p_2, p_3$ both inside the absorbing phase transition plane, $p_1=0, p_{2/3} \neq 0$, and if we leave it $p_1 \neq 0$, as these deformations correspond to continuously tuning the states $\ket{i_{j_1\cdots j_6}}$ the Hamiltonian projects to. Discontinuities can arise when multiple $p_i$ are tuned to zero, as different paths in the parameter space can lead to different states $B_\vee$ projects to. For our purposes, this is not relevant; in particular, the Hamiltonian remains continuous across the DP phase transition line.

Let us consider again $p_1=0$, for which states with different defect number sectors are not coupled. This can be directly expressed as a $U(1)$ conservation law; defining a local defect operator $n_{D,\vee}$ in the space $i_1 = i_2$ as
\begin{equation}
    n_{D,\vee} = \frac{1}{16}(1-Z_{i_1})(1-Z_{i_2})(1+Z_{j_1})(1+Z_{j_2}),
\end{equation}
the total number of defects $N_D = \sum_\vee n_{D,\vee}$ is conserved as $[H_{\text{DK}}, N_D] =0$. The Hilbert space thus splits up into sectors labeled by the eigenvalues of $N_D$ (see bottom of \fig{fig:ParentH} for a pictorial representation of the parent Hamiltonian). The Hamiltonian connects any state with a non-zero defect number to a state where all defects are isolated $\ket{j_1\cdots j_6} = \ket{0\cdots 0}\; \forall \;\vee$; this state violates the associated projector $B_\vee \ket{0\cdots 0}\ket{11} = \ket{0\cdots 0}\ket{11}$ and cannot be in the ground state manifold. Thus, all ground states lie in the $N_D = 0$ sector. The vacuum state $\ket{\text{vac}} = \ket{00\cdots 0}$ trivially is a ground state as it is annihilated by all projectors; it cannot be connected to any other state with $N_D = 0$ through a local operation. The sector thus splits up further into the single state $\ket{\text{vac}}$ and all remaining states, including the second ground state $\ket{\text{GS}}$ (\ref{eq:GroundStateNonTriv}).

\section{Appendix B: Subsystem Negativity for $p_1 = 0, p_{2/3} \ll 1$}
In this section, we derive the negativity of the two-subspace reduced density matrix stated in the main text. We consider a system with periodic boundary conditions with linear system sizes $L_x$ and $L_y$, where $x/y$ corresponds to the spatial/temporal direction of the (1+1)D stochastic automaton. As discussed in the main text, deep in the absorbing phase $p_1 = 0, p_{2/3} \ll 1$, the non-trivial ground state $\ket{\text{GS}}$ approaches the limit of a superposition of a single string of active sites on top of the vacuum which wraps around the $y$ direction once. This ``fixed point" can be expressed as
\begin{equation}
    \ket{\psi_{\text{1string}}} \propto \sum_{\text{s around $L_y$}} \left( \prod_{\vee \in s} X_{i_1}X_{i_2} \right) \ket{00 \cdots 0},
\end{equation}
where $s$ represents a closed path of length $L_y$ in which each value for $y$ appears precisely once.

We choose two rectangular subregions $A$ and $B$ of the total system spanning the entire $y$ direction, with finite extent $\ell > L_y$ in $x$ direction. To probe the entanglement between the two regions, we express their reduced density matrix $\rho_{AB}$, obtained by tracing out the remaining system, in the general form
\begin{equation}
    \rho_{AB} = \alpha_0 \ket{0_A}\ket{0_B}\bra{0_A}\bra{0_B} + \alpha_\Phi \ket{\Phi_s}\bra{\Phi_s} + \sum_i \alpha_{\Psi_i} \ket{\Psi_i}\bra{\Psi_i}
    \label{eq:ReducedDensityMatrix}
\end{equation}
with $\alpha_0 + \alpha_\Phi + \sum_i \alpha_{\Psi_i} = 1$. The first contribution comes from all configurations where the single string is outside $A \cup B$, while for the second contribution it lies inside $A \cup B$; the state $\ket{\Phi_s} =\frac{1}{\sqrt{2}} (\ket{0_A} \ket{s_B} + \ket{s_A} \ket{0_B})$, where $\ket{s_{A/B}}$ is the superposition of all string configurations in $A/B$, is a Bell state of the string. All other contributions $\ket{\Psi_i}$ come from cases where the strings lies partially inside and partially outside $A \cup B$; they factorize into states in $A$ and $B$, $\ket{\Psi_i} = \ket{i_A} \ket{i_B}$.

A computable measure of entanglement for bipartite mixed density matrices is the negativity $\mathcal{N}$~\cite{horodecki_entanglement2009, vidal_computable2002}. Any bipartite density matrix $\rho_{AB}$ with $\mathcal{N}(\rho_{AB}) > 0$ is entangled. Introducing the partial transpose $\Gamma_B$ in the subsystem $B$, it is defined as
\begin{equation}
    \mathcal{N}(\rho_{AB}) = \frac{\vert\vert\rho_{AB}^{\Gamma_B}\vert\vert_1 -1}{2} = \sum_{i} \frac{\vert \lambda_i \vert - \lambda_i}{2},
    \label{eq:Negativity}
\end{equation}
where $\vert\vert \cdot\vert\vert_1$ is the trace norm and $\lambda_i$ are the eigenvalues of $\rho_{AB}^{\Gamma_B}$. $\mathcal{N}$ is the absolute sum of its negative eigenvalues and quantifies its violation of the positive partial trace criterion~\cite{peres_separability1996, horodecki2001separability}.

The only part of $\rho_{AB}$ Eq.~(\ref{eq:ReducedDensityMatrix}) which the partial transpose $\Gamma_B$ acts on non-trivially is the Bell state,
\begin{widetext}
    \begin{align}
        \Bigl(\ket{\Phi_s} \bra{\Phi_s}\Bigr)^{\Gamma_B} &= \frac{1}{2} \Bigl(\ket{0_A} \ket{s_B} \bra{0_A} \bra{s_B} + \ket{s_A} \ket{0_B} \bra{0_A} \bra{s_B} + \ket{0_A} \ket{s_B} \bra{s_A} \bra{0_B} + \ket{s_A} \ket{0_B} \bra{s_A}\bra{0_B} \Bigr)^{\Gamma_B} \nonumber \\ & =\frac{1}{2} \Bigl(\ket{0_A} \ket{s_B} \bra{0_A} \bra{s_B} + \ket{s_A} \ket{s_B} \bra{0_A} \bra{0_B} + \ket{0_A} \ket{0_B} \bra{s_A} \bra{s_B} + \ket{s_A} \ket{0_B} \bra{s_A}\bra{0_B} \Bigr).
    \end{align}
\end{widetext}
The four-dimensional space spanned by the states $\ket{0_A} \ket{0_B},\ket{s_A} \ket{0_B}, \ket{0_A} \ket{s_B}, \ket{s_A} \ket{s_B}$ is not connected to any of the remaining states $\ket{\Psi_i}$ by $\rho_{AB}^{\Gamma_B}$, as they by definition include a partial string in $A$ or $B$. Treating this sector separately, it splits into two $1 \times 1$ sectors of single eigenvalues $\ket{s_A} \ket{0_B}$ and $\ket{0_A} \ket{s_B}$ with positive eigenvalue $\frac{\alpha_\Phi}{2}$ and a $2 \times 2$ sector spanned by $\ket{0_A} \ket{0_B}$ and $\ket{s_A} \ket{s_B}$ of the form $\begin{pmatrix} \alpha_0 & \frac{\alpha_\Phi}{2} \\ \frac{\alpha_\Phi}{2} & 0 \end{pmatrix}$, which for $\alpha_\Phi \neq 0$ has a negative eigenvalue $\lambda_0 = \frac{\alpha_0}{2}\left[1-\sqrt{1+\left(\frac{\alpha_\Phi}{\alpha_0}\right)^2}\;\right]$. Furthermore, as $\Bigl(\ket{\Psi_i}\bra{\Psi_i}\Bigr)^{\Gamma_B} = \ket{\Psi_i}\bra{\Psi_i}$, the remaining eigenstates of $\rho_{AB}^{\Gamma_B}$ are the $\ket{\Psi_i}$ with positive eigenvalues $\alpha_{\Psi_i}$, so $\lambda_0$ is the only negative eigenvalue and $\mathcal{N}(\rho_{AB}) = \vert\lambda_0 \vert $.

We find a lower bound for this expression in terms of the linear system size $L_x$ by means of a lower bound for the coefficient $\alpha_\Phi$. As the state is translationally invariant in $x$ direction and has precisely one $\ket 1$ state per layer in $y$ direction, the probability of finding this $\ket 1$ at some $x$ coordinate in a given layer $y_0$ is $\frac{1}{L_x}$. As $A$ and $B$ both have an extent in $x$ direction $\ell > L_y$, the entire string has to be in this subsystem if the $\ket 1$ is in its middle at $y_0$. We therefore find $(\bra{0_A}\bra{s_B})\ket{\psi_{\text{1String}}} = (\bra{s_A}\bra{0_B})\ket{\psi_{\text{1String}}} \geq \frac{1}{\sqrt{L_x}}$ and thus $\alpha_\Phi \geq \frac{2}{L_x}$. Furthermore, as the trace of $\rho_{AB}$ is one, $\alpha_0 \leq \frac{L_x -2}{L_x}$. From this we obtain the lower bound on the negativity

\begin{align}
    \mathcal{N}(\rho_{AB}) &= \frac{\alpha_0}{2}\left[\sqrt{1+\left(\frac{\alpha_\Phi}{\alpha_0}\right)^2} -1\right] \geq \frac{\alpha_0}{2}\left[\sqrt{1+\left(\frac{2}{L_x\alpha_0}\right)^2} -1\right] \nonumber \\ & \geq \frac{L_x -2}{2L_x}\left[\sqrt{1+\left(\frac{2}{L_x -2}\right)^2} -1\right] = \frac{1}{L_x(L_x -2)} + \mathcal{O}(\frac{1}{L_x^4}).
\end{align}
This result can be compared with the corresponding expression for the $W$ state of $L$ qubits defined as $\ket W = \frac{1}{L} \left( \sum_i^L X_i \right) \ket{00 \cdots 0}$. The reduced density matrix of two qubits is $\rho_{AB} = \frac{L-2}{L} \ket{00}\bra{00} + \frac{2}{L}\ket{\Phi^+}\bra{\Phi^+}$, where $\ket{\Phi^+} = \frac{1}{\sqrt{2}}(\ket{01} + \ket{10})$ is a Bell state, leading to the negativity saturating the lower bound $\mathcal{N}(\rho_{AB}) = \frac{L-2}{2L}\left[\sqrt{1+\left(\frac{2}{L -2}\right)^2} -1\right] = \frac{1}{L(L -2)} + \mathcal{O}(\frac{1}{L^4})$. In both cases, the distance $d$ does not appear in the expression; even if the regions/qubits $A$ and $B$ are far apart, the pairwise entanglement remains constant. In this sense, the state $\ket{\psi_{\text{1string}}}$ exhibits a similar entanglement pattern as $\ket W$, which remains stable in the absorbing phase below the critical probability, as the single cluster simply broadens on the scale of the perpendicular correlation length $\xi_\perp$; the main feature of a delocalized finite-size object remains present.

\let\oldaddcontentsline\addcontentsline
\renewcommand{\addcontentsline}[3]{}
\bibliography{references}
\let\addcontentsline\oldaddcontentsline

\newpage
\leavevmode \newpage

\setcounter{equation}{0}
\setcounter{page}{1}
\setcounter{figure}{0}
\renewcommand{\thepage}{S\arabic{page}}  
\renewcommand{\thefigure}{S\arabic{figure}}
\renewcommand{\theequation}{S\arabic{equation}}
\onecolumngrid
\begin{center}
\textbf{Supplemental Material:}\\
\textbf{\papertitle}\\ \vspace{10pt}
\vspace{6pt}
Julian Boesl$^{1,2}$, Frank Pollmann$^{1,2}$ and Michael Knap$^{1,2}$ \\ \vspace{6pt}
$^1$\textit{\small{Technical University of Munich, TUM School of Natural Sciences, Physics Department, 85748 Garching, Germany}} \\
$^2$\textit{\small{Munich Center for Quantum Science and Technology (MCQST), Schellingstr. 4, 80799 M{\"u}nchen, Germany}} \\
\vspace{10pt}
\end{center}

\maketitle

\twocolumngrid

\section{Quantum state from Kasteleyn model}
We discuss quantum states from mappings of other classical statistical models which have similar properties on first sight, but as we will show nevertheless do not realize the same physics. While these parametrized tensor network states can also undergo a macroscopic change of the wavefunction, there is no criticality at play due to additional conservation laws of the parent Hamiltonians.

The thermodynamic limit of the state $\ket{\text{DK}}$ in the $p_1 = 0$ plane differs in the two phases; while in the absorbing phase $\lim_{L_x,L_y \rightarrow \infty}\ket{\text{DK}} = \ket{\text{vac}}$ as all clusters are suppressed in the length $L_y$, in the active phase the state remains distinct from the vacuum state due to the existence of a non-trivial steady state. This survival of system-size spanning fluctuations in the thermodynamic limit is reminiscent of the classical Kasteleyn model~\cite{kasteleyn1963dimer, powell_quantumkasteleyn2022}. This model considers hardcore dimers on edges of a honeycomb lattice with periodic boundary conditions. Each site of the lattice is touched by precisely one dimer, leading to close-packed configurations. There are three directions for the edges; we designate one of them to be in $y$ direction. The energy of a configuration $\alpha$ is given by $E_\alpha = -VN_y$, where $N_y$ is the number of dimers on edges parallel to the $y$ direction. The partition function of the model at inverse temperature $\beta$ is given by $Z_K = \sum_\alpha e^{-\beta E_\alpha}$.

The state with minimal energy is given by a configuration where all dimers live on $y$ edges. Due to the hard-core constraint, the energy of the lowest excited states scales as $\propto L_y$, as the dimers have to be rearranged across the entire $y$ direction. When compared to the lowest energy state as a reference configuration, excited states can thus be described by strings winding around the $y$ direction, each string incurring an energy cost on the scale of the linear system size $L_y$. This dependence on $L_y$ leads to an atypical phase transition of the Kasteleyn model in the thermodynamic limit $L_x, L_y \rightarrow \infty$: Below a critical value $V\beta_c = \ln 2$, all excited states remain suppressed and the partition sum is dominated by the zero-energy configuration. Above this critical temperature, the entropic gain from different string positions overtakes the energy penalty, leading to a finite density of strings even in the thermodynamic limit.

\begin{figure}
\centering
    \includegraphics[width=.75\columnwidth]{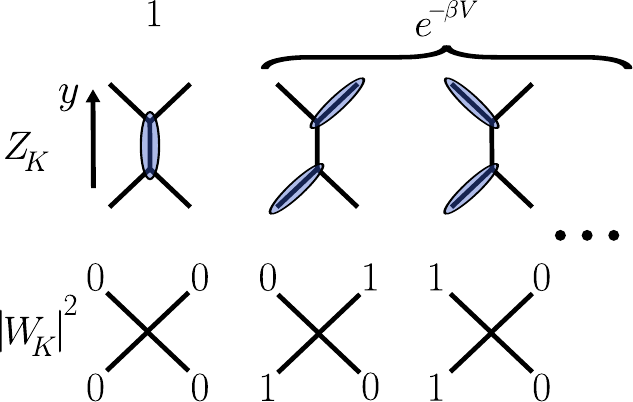}
    \caption{\textbf{Mapping the Kasteleyn model to a TNS.} In the classical Kasteleyn model on the honeycomb lattice, each vertex is covered by precisely one dimer; choosing one of the edge types to lie in $y$ direction, the system prefers the dimers to sit on these edges through a local potential $V$, penalizing all other local configurations with a Boltzmann factor $e^{-\beta V}$ in the partition sum $Z_K$. As the presence or absence of a dimer is determined by the adjacent edges, we can express the system on a square lattice. Configurations $\alpha$ differing from the lowest-energy configuration can be represented by strings of $1$'s winding around the $y$ directions. The associated quantum state is represented by a matrix $W_K$ with virtual legs, where virtual and physical legs on the edges are locked; for every state $\ket \alpha$, the overlap is then given by the Boltzmann weight $\vert \langle \alpha \vert\psi_K\rangle\vert^2 = \frac{e^{-\beta E_\alpha}}{Z_K}$.
    }
    \label{fig:Kasteleyn}
\end{figure}

As with other classical lattice models, the Kasteleyn model can be mapped to a quantum state~\cite{verstraete_criticality2006}. First, for simplicity we note that the honeycomb lattice can be mapped to a square lattice: Whether an edge in $y$ direction is occupied by a dimer is uniquely determined by the four adjacent edges; its information is thus redundant. Similar to the Domany-Kinzel automaton, we put a physical qubit on each edge with $\ket{0/1}$ signifying the absence/presence of a dimer on this leg (see \fig{fig:Kasteleyn} for a graphical representation of the mapping). We can thus define a tensor $(T_K)^{ij}_{abcd} = \sum_{a^\prime b^\prime} \delta_{aa^\prime}^i \delta_{bb^\prime}^j (W_K)_{{(a^\prime b^\prime )(cd)}}$, which we further decompose by pulling out the ``plumbing" tensors $\delta_{aa^\prime}^i$ defined as $\delta_{ab}^\sigma = 1$ if $a = b = \sigma$ and zero otherwise~\cite{liu_simulating2024, boesl2025skeleton}. The information about the state is thus contained entirely in the matrix $W_K$, which is of the form
\begin{equation}
        W_{K}(g) =  \begin{pNiceMatrix}[first-row,last-col] \ket{00} & \ket{01} & \ket{10} & \ket{110} & \\ 1 & 0 & 0 & 0 & \; \ket{00} \\ 0 & g & g & 0 & \; \ket{01} \\ 0 & g & g & 0 & \; \ket{10} \\ 0 & 0 & 0& 0 & \; \ket{11} \end{pNiceMatrix},
        \label{eq:Kasteleyn}
\end{equation}
where $g = e^{-\frac{\beta V}{2}}$ is the square root of the local Boltzmann factor when there is no dimer on the contracted $y$ edge. The contracted tensor network state is $\ket{\psi_K(g)} = \frac{1}{Z_K}\sum_\alpha e^{-\frac{E_\alpha}{2}} \ket \alpha$. Similar to the DK state, this state has two different regimes when approaching the thermodynamic limit, as $\lim_{L_x,L_y \rightarrow \infty}\ket{\psi_K(g)} = \ket{\text{vac}}$ if $g \leq \frac{1}{\sqrt{2}}$, while $\lim_{L_x,L_y \rightarrow \infty}\vert\langle\text{vac}\ket{\psi_K(g)}\vert^2 < 1$ for $g > \frac{1}{\sqrt{2}}$.

However, on the level of the quantum state the origin of this abrupt change is different from the transition discussed in the main text. To understand this, we may construct a parent Hamiltonian $H_K$ of $\ket{\psi_K(g)}$. We can again find a Hamiltonian built of frustration-free local projectors; they should penalize all configurations with open or branching strings in their support, as these are not allowed by the tensor in \eq{eq:Kasteleyn}. Furthermore, the Hamiltonian can only couple states with the same number of strings around the $y$ direction, as a change in string number would amount to a non-local operation; the smallest non-trivial projector satisfying this is thus a plaquette operator which hops a string passing the plaquette on the left to one passing it on the right and vice versa, if possible. However, even if we allowed for bigger support of the projectors, the Hamiltonian $H_K$ would always conserve the number of strings $N_{\text{strings}}$ in the system. If we thus consider generalizations of the state $\ket {\psi_{1\text{string}}}$ labeled $\ket {\psi_{N\text{strings}}}$ which are equal weight superpositions of all configurations of $N$ strings which do not touch each other, we see that they are also ground states of $H_K$ alongside $\ket{\text{vac}}$ and $\ket{\psi_{1\text{string}}}$. Importantly, this Hamiltonian does not depend on the parameter $g$; the state $\ket{\psi_K(g)}$ thus simply represents a parametrized state in the ground state manifold, where the overlap with each basis state is controlled by $g$. For $g$ below the critical value $g_c = \frac{1}{\sqrt{2}}$, $\lim_{L_x,L_y \rightarrow \infty}\vert\langle\psi_\text{1string}\ket{\psi_K(g)}\vert^2 = \lim_{L_x,L_y \rightarrow \infty}\vert\langle\psi_{N\text{strings}}\ket{\psi_K(g)}\vert^2 = 0$, while above this threshold the overlap remains finite.

The state $\ket{\psi_K(g)}$ falls under a more general class of quantum states which are superpositions of strings configurations wrapping around one direction; for example, the matrix $W_K$ is a special case of
\begin{equation}
    W(\{ g \}) =  \begin{pmatrix} g_1 & 0 & 0 & 0 \\ 0 & g_2 & g_3 & 0 \\ 0 & g_4 & g_6 & 0 \\ 0 & 0 & 0& g_7 \end{pmatrix}.
\end{equation}
Another special case is given by the choice $g_1 = g_7 = 1, g_2 = g_3 = g_4 = g_5 = \frac{1}{\sqrt{2}}$ which can be mapped to the six-vertex model and arises as the critical point between two phases with the topological order of the toric code, distinguished by different symmetry fractionalization patterns~\cite{liu_simulating2024}. All these states share the propertry that the number of $\ket{1}$ states in each layer in $y$ direction is conserved. If $g_1 \neq 0$ and $g_2 = g_3 = g_4 = g_5$, the state $\ket{\psi_{\text{1string}}}$ is a ground state of the parent Hamiltonian. Nevertheless, the Hilbert space of these parent Hamiltonians is not connected enough to realize the transition discussed in the main text, as there is no way to increase width of the string. For the DK state on the other hand, which is represented by a matrix
\begin{equation}
    W_{\text{DK}}(p_1, p_2, p_3) =  \begin{pmatrix} \sqrt{1-p_1} & \sqrt{1-p_2} & \sqrt{1-p_2} & \sqrt{1-p_3} \\ 0 & 0 & 0 & 0 \\ 0 & 0 & 0 & 0 \\ \sqrt{p_1} & \sqrt{p_2} & \sqrt{p_2}& \sqrt{p_3} \end{pmatrix},
\end{equation}
$N_{\text{strings}}$ is not a well-defined quantity as clusters can branch off. This higher connectivity of the $N_D = 0$ sector of $H_{\text{DK}}$ (except for the state $\ket{\text{vac}}$) allows the entanglement pattern transition to take place for the DK state but not for the Kasteleyn state.

\section{3D Quantum state from 2D bond directed percolation}
In the main text, we have discussed the Domany-Kinzel automaton as it is a particularly well-understood example of a low-dimensional stochastic circuit in the directed percolation universality class. However, the mapping can be applied to a variety of probabilistic lattice automata with absorbing phase transitions in different dimensions and different universality classes. The corresponding quantum systems can thus live on different lattices and can have different local Hilbert spaces, while the transition between different classes of states can be understood similarly in terms of the entanglement pattern. The critical exponents observed at the transition between the two classes depend on the universality class and the dimension of the system.

\begin{figure}
\centering
    \includegraphics[width=.99\columnwidth]{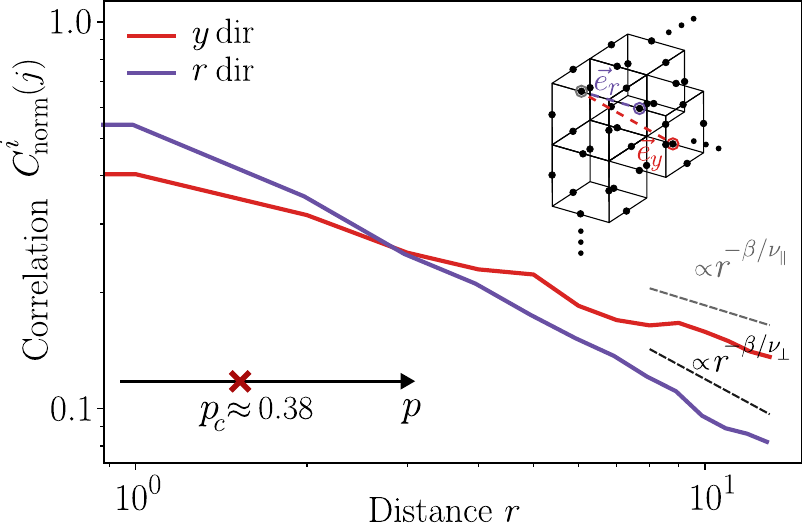}
    \caption{\textbf{Normalized correlations of a 3D critical isoTNS.} The normalized correlations $C^i_{\text{norm}}(j)$ in two directions, evaluated in an open boundary system with fully active boundaries, for the bond directed percolation process on a cubic lattice, corresponding to a 3D isoTNS. We show results at the approximate critical point $p_c \approx 0.38$. All directions feature algebraic scaling, with the exponent depending on the direction, $\beta/\nu_\parallel$ in $y$ direction (red), $\beta/\nu_\perp$ in all other directions, here represented by a lattice direction $\vec{e}_r$ (purple). The linear system size is $L = 400$.
    }
    \label{fig:3bDP}
\end{figure}

In this section, we provide another simple example of such a transition. The state lives on a three-dimensional cubic lattice and is represented by a tensor $T_{2\text{bDP}}$ on a vertex of the lattice, with six virtual degrees of freedom $a,b,c,d,e,f$ of bond dimension $\chi = 2$ in each direction and three physical degrees $j,k,l$ of freedom on an edge into each direction; the physical legs are qubits which are locked to the virtual degree of the corresponding edge. As in the main text, the values of the tensor are determined by a stochastic update rule, as $(T_{2\text{bDP}})^{jkl}_{abcdef} = \sqrt{P(i\vert jkl)} \delta_{a,j} \delta_{b,k} \delta_{c,l} \delta_{d,i}\delta_{e,i}\delta_{f,i}$. Here, the associated (2+1)D stochastic automaton realizes bond directed percolation, i.e. every active site in the set $j,k,l$ leads with probability $p$ to an active site $i$, with the fully empty state as an absorbing state. In the Domany-Kinzel automaton, the one-dimensional version of bond directed percolation is realized on the line $p_3 = p_2(2-p_2)$ at $p_1 = 0$ with a critical point at $p_2 \approx 0.645$~\cite{hinrichsen_nonequilibrium2000}. On our lattice, the higher-dimensional variant is given by the update rule
\begin{align}
    &P(1\vert000) = 0, \;P(1\vert001) = P(1\vert010) = P(1\vert100)  = p, \nonumber\\ &P(1\vert011) = P(1\vert101) = P(1\vert110)  = 1-(1- p)^2, \nonumber\\ &P(1\vert111) = 1- (1-p)^3
    \label{eq:bDPDefinition}
\end{align}
The conservation of probability in the circuit guarantees that the tensor $T_{{2\text{bDP}}}$ is an isoTNS. Similar to its lower-dimensional version, we can expect this two-dimensional version of bond directed percolation to undergo a phase transition from an absorbing to an active phase at some critical probability $p_c$, the process at $p_c$ exhibiting critical scaling with exponents $\beta, \nu_\parallel, \nu_\perp$ given by the 2D directed percolation universality class. We evaluate the normalized correlation functions  $C^i_\text{norm}(j)$ close to the numerical estimate of the critical probability $p_c \approx 0.38$~\cite{grassberger_directed_1989}, where they exhibit the expected critical scaling $C^i_\text{norm}(j) \sim \vert i-j\vert^{-\beta/\nu_\parallel}$ in the time-like diagonal direction $y$ and $\sim \vert i-j\vert^{-\beta/\nu_\perp}$ in all other directions, with $\beta/\nu_\parallel \approx 0.451$ and $\beta/\nu_\perp \approx 0.796$ as estimated in the literature~\cite{ odor_universality2004} (See \fig{fig:3bDP}). $T_{{2\text{bDP}}}$ thus represents a 3D isoTNS with critical correlations in all directions.

The interpretation of the classes of states corresponding to the absorbing and active phase of the automaton for periodic boundary conditions is analogous to the 2D case, with the ground state manifold of the parent Hamiltonian always including the trivial state $\ket{\text{vac}}$, which is necessary to stabilize the transition, and a second ground state $\ket{\text{GS}}$ which represents a single delocalized cluster in the absorbing phase and a trivially entangled state in the active phase.

In a similar fashion, generalized Domany-Kinzel automata with multiple absorbing states can also be mapped to quantum states. The local Hilbert space dimension on the edges increases to accommodate the higher number of states each site can be in; the multiple absorbing states lead to additional exact eigenstates in the $N_D = 0$ sector of the quantum parent Hamiltonian, while the presence of a phase transition and the universality class depend on the number of absorbing states and the dimension; for two absorbing states in one dimension, the so-called DP2 universality class is realized~\cite{hinrichsen_stochastic1997}. Studying the mapping of such and other classical stochastic automata to quantum states is an exciting future research direction. 

\end{document}